\documentclass[]{aa}  
\usepackage{graphicx}
\usepackage{longtable}
\usepackage{txfonts}

\usepackage{natbib}
\input{psfig}

\begin{document}

     \title{Modelling solar-like variability for the detection of Earth-like planetary transits}
\subtitle{I. Performance of the three-spot modelling and harmonic function fitting }

    \titlerunning{Solar-like activity and planetary transits}
    \authorrunning{A. S. Bonomo \& A. F. Lanza}


   \author{A.~S.~Bonomo\inst{1,2}, A.~F.~Lanza\inst{2}}

   \offprints{A. S. Bonomo}

   \institute{Dipartimento di Fisica e Astronomia, Sezione Astrofisica, Universit\`a degli Studi di Catania, \and    
             INAF-Osservatorio Astrofisico di Catania, Via S. Sofia, 78 -- 95123 Catania, Italy \\ 
              \email{aldo.bonomo@oact.inaf.it, nuccio.lanza@oact.inaf.it}    
             }

   \date{Received 11 October 2007 ; accepted 20 February 2008 }

    \abstract
{}{We present a comparison of two methods of fitting solar-like variability to increase 
the efficiency of detection of Earth-like planetary transits across the disk 
of a Sun-like star. { One of them is the harmonic fitting method that coupled with the BLS detection algorithm  demonstrated the best performance during the first CoRoT blind test.}}{{ We apply a Monte Carlo approach by simulating a large number of light 
curves of duration 150 days  for different values of  planetary radius, orbital period, epoch of the first transit, and standard deviation of the photon shot noise}. Stellar variability is assumed in all the cases to be given by the Total Solar Irradiance variations as observed close to the maximum of solar cycle 23. { After fitting solar variability, transits are searched for by means of the BLS algorithm.}}{We find that a model based on three point-like active regions is better suited than a best fit with a linear combination of 200 harmonic functions to reduce the impact of stellar microvariability provided that the standard deviation of the noise is $2-4$ times larger than the central depth of the transits. On the other hand, the 200-harmonic fit is better when the standard deviation of the noise is comparable to the transit depth.}{Our results show the advantage of a model including a simple but physically motivated  treatment of stellar microvariability for the detection of planetary transits when the standard deviation of the photon shot noise is greater than the transit depth and stellar variability is analogous to solar irradiance variations.}
\keywords{planetary systems -- methods: data analysis -- techniques: photometric -- stars: activity -- stars: late-type}

   \maketitle
%

\section{Introduction}

Since the pioneering work by \cite{MajorQueloz95}, about 270 extrasolar planets
have  been discovered, mostly through the periodic oscillations of the radial
velocity of their parent stars\footnote{See http://exoplanet.eu/}. About  35 
show transits across the disk of their stars allowing us to remove 
degeneracy between orbital inclination and amplitude of the radial velocity curve,
thus providing information on planet's mass and radius. Transit searches  from the ground are limited to
giant planets because their photometric precision cannot be pushed beyond the millimagnitude limit 
due to the scintillation of the Earth atmosphere. 
This limitation can be overcome by space-borne telescopes. The space missions
CoRoT and Kepler have been specifically designed for searching planetary transits and  
can reach the photometric precision  to detect  transits of terrestrial planets
in front of late-type main-sequence stars \citep{Baglin03,Moutouetal05,Boruckietal04}.

The main source of noise in the detection of terrestrial planet transits at the level of precision 
of space-borne photometry is the intrinsic variability of the parent star. For late-type stars 
having an outer convection zone,
magnetic activity is the main source of variability and can severely limit
the capability of detecting transits even in relatively inactive stars. 
In the case of the Sun, the transit of a large sunspot group across the
disk produces a dip of the solar irradiance of $2000-2500$ parts per million (ppm),
whereas the central transit of an earth-sized planet produces a flux decrease
of only 84 ppm \citep[][]{Jenkins02}. During phases of minimum activity, even when
no sizeable spot group is detected, the solar flux is modulated by faculae, with 
amplitudes usually exceeding 100-200 ppm. 

In principle, the extraction of the transit signal from a light curve contaminated by  
stellar microvariability is  possible thanks to the different timescales
characterizing transit duration  and stellar variability. Specifically, the former 
ranges from  a few hours to about one day, in the case of 
terrestrial planets with orbital periods up to about one year around a solar-like star, while 
the latter  is dominated by the rotational modulation of the largest active regions and
their evolution, taking place on time scales from weeks to months in the case of the 
Sun \citep[cf., e.g., ][ and references therein]{Jenkins02,Lanzaetal03, Lanzaetal07}.

Several methods
have been proposed to reduce the impact of solar-like variability on transit detection, 
starting from simple high-pass filters \citep{Defayetal01} to optimal
filters \citep{Jenkinsetal02,Carpanoetal03}. Those methods work well only for continuous,
evenly sampled timeseries because gaps or irregular sampling introduce  aliasing problems. Moreover, 
optimal filters need a sufficiently long timeseries to estimate the power spectrum of 
the intrinsic stellar variability from the data themselves in order to construct the filter. 
For a slowly rotating star as the Sun, this implies a time series of at least 
two or three months, given that most of the variability is on rotational timescales.
Recently \citet{AigrainIrwin04} have proposed methods to overcome the difficulties due to
irregular sampling or gaps, based on 
a least square fit of individual sine and cosine components or on the application of iterative non-linear
filters. 

The first CoRoT blind test \citep{Moutouetal05} represented a testbed for several
different methods to treat solar-like variability and showed the potentiality of those methods
in combination with suitable transit finding algorithms to detect transits in a quite large
set of simulated light curves including also gaps and instrumental effects.  The best method
to remove stellar microvariability proved to be a best fit based on the use of 200 sine and cosine
harmonic functions coupled with the BLS transit finder algorithm \citep{Kovacsetal02}, proposed by the Geneva team. 

In the present paper, we propose a different method to treat stellar variability,  based on our 
model of the Total Solar Irradiance (hereinafter TSI)
variations \citep{Lanzaetal03,Lanzaetal07}. { Previous models do not exploit  
knowledge of the cause of the variation, that is the intrinsic evolution and the 
rotational modulation of the visibility of active regions, while our model includes such effects.} 
The residuals obtained after subtracting our model have a maximum amplitude 
of  100-150 ppm and an almost Gaussian distribution 
\citep[cf. ][ and Sect.~\ref{3spots}]{Lanzaetal03}, suggesting us to apply  algorithms developed for the 
detection of planetary transits in the presence of a Gaussian white noise. The performance of 
the proposed approach will be  compared with that of the 200-harmonic best-fit algorithm
of the CoRoT blind test to establish advantages and  limitations of
the new method. Moreover,  other methods will be briefly discussed in
Sect.~\ref{discuss}, deferring a detailed comparison to successive papers. 
 
\section{Light curve simulations}

To assess the performance of the proposed method, we shall analyse an
extensive set of simulated light curves. Each light curve includes 
flux variations produced by Sun-like activity, planetary transits and photon shot noise, respectively.

\begin{figure*}[t]
\includegraphics[width=6cm,height=16cm,angle=90]{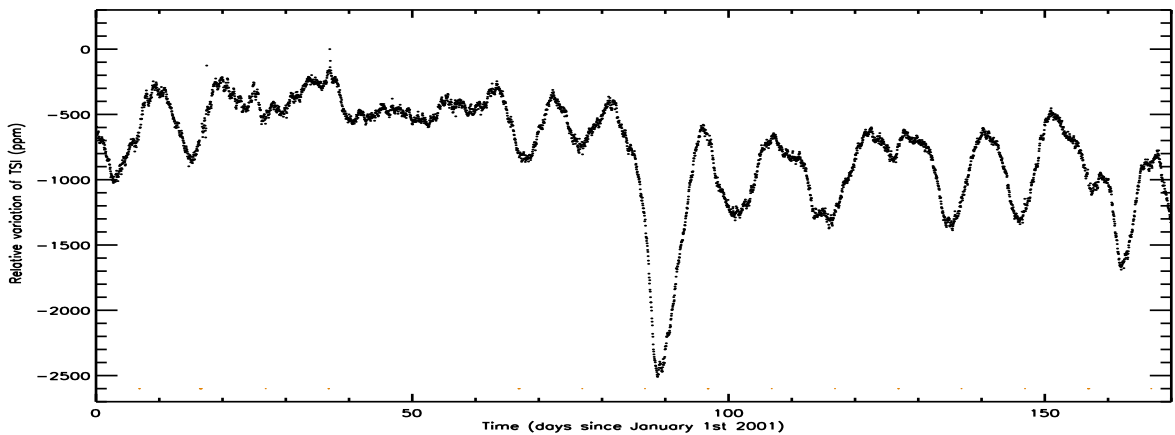} 
\caption{The Total Solar Irradiance time series (level 2.0 data of the VIRGO experiment) used to simulate
Sun-like microvariability in the present investigation. Small horizontal red dashes at the bottom of the plot
mark  gaps in the time series. } 
\label{tsi_plot}
\end{figure*}

{ To simulate the effects of Sun-like activity, we choose a subset of the TSI variation of  duration 170 days ranging from 
January 1$^{\rm st}$ to June 19$^{\rm th}$ 2001 as observed by the radiometers of the
VIRGO experiment on board of the SoHO satellite 
\citep[][ see Fig.~\ref{tsi_plot}]{Frolichetal95,Frolichetal97}}. Specifically,
we use the 2.0-level TSI time series with a time sampling of one value per hour, accessible
from the VIRGO web site at the Physikalisch-Meteorologisches Observatorium
Davos -- World Radiation Center\footnote{http://www.pmodwrc.ch/pmod.php?topic=tsi
/virgo/proj\_space\_virgo\#VIRGO\_Radiometry}. The standard deviation of the
hourly TSI data is 20 ppm. 
Short quasi-periodic gaps,  not exceeding a few hours, are present in the TSI time series and are retained in
our simulations as representative of the gaps expected for a space-borne
photometric experiment. Their total duration is  36 hours, i.e., 0.9 per cent of the time series 
(see Fig.~\ref{tsi_plot}). 

The considered time series falls close to the maximum of activity cycle 23 and shows flux 
variations up to 2600 ppm due to the transits of large active regions across the solar disk. 
Therefore, it is well suited to study the impact of the largest Sun-like variability on planetary
transit detection.  
Note  that the level of activity of the Sun at the maximum of the 11-yr cycle is close to the median 
activity level of field dwarf stars of spectral types F and G, as found by \cite{Schmitt97} by comparing their
coronal X-ray fluxes.  

We prefer to use real solar irradiance data instead of simulating stellar 
microvariability with the approaches proposed by, e.g., \citet{Aigrainetal04} or \citet{Lanzaetal06}. 
Actually, those models are based on extrapolations of the solar variability characteristics to the case of
more active stars and cannot be tested observationally yet.  

{ To simulate stellar variability for intervals of 150 days, such as those of the long runs of the space experiment CoRoT, we extract 150-d sections of the above TSI time series choosing starting points at random between 0 and 20 days. In such a way, we get different realizations of Sun-like activity  reducing possible systematic effects associated with the  variations contained in a fixed TSI subset (cf. Sect.~\ref{3spots})}. 

The light curves of planetary transits are computed by means of the IDL procedure
UTM \citep{Deeg99}. 
We consider circular planetary orbits with periods of 5, 10, 25 and 50 days, and an inclination of
$90^{\circ}$, i.e., the planet transits across the centre of the disk of the star. 
The radius of the planet is varied between 1.0 and 2.0 R$_{\oplus}$ in steps of $0.25$ R$_{\oplus}$. 
A radius of 1 R$_{\odot}$ and a mass of 1 M$_{\odot}$ are assumed for the star and 
a quadratic limb-darkening law is adopted for the stellar photosphere. 
Therefore, the duration of the transits ranges from 3.11 hours for an orbital period of 5 days up to 6.67 hours for
a period of 50 days. 

{ In such a way, a total of 20 different transit light curves has been simulated. Finally, we shift the initial transit to a random phase between 0.0 and 1.0, taken from a  uniform distribution, and add stellar variability and photon 
shot noise  assuming that  Poisson statistics can be well approximated by Gaussian statistics
at the photon count levels characteristic of planetary transit searches. 
 We consider four levels
of noise, i.e., with standard deviations of 100, 200, 300 and 1000 ppm,  which  approximately 
correspond to the noise levels reached in one hour integration time for  G2V stars of   magnitudes $V = 12$, 13, 14 and 16 as observed by CoRoT, respectively\footnote{\mbox{http://corotsol.obspm.fr/web-instrum/payload.param/index.html}}. 
One hundred independent white Gaussian noise realizations are computed 
for each value of the standard deviation and added to each of the noiseless light curves giving a total of 
8,000 light curves with transits}. 

In a search for planetary transits it is important to estimate the number of false alarms produced by 
those random combinations of the noise that may approximate a transit feature 
\citep[see ][ and Sect.~\ref{alpha_deter}]{Doyleetal00}.
{ We determine the frequency
of false alarms by analysing transitless light curves obtained by combining solar irradiance variations with
100 random sequences for each of the above noise standard deviations, for a total of 400 light curves. } 

\section{Analysis of  simulated light curves}

The simulated light curves are analysed in two steps. First we fit stellar microvariability 
by applying our model based on three active regions 
\citep[hereinafter 3-spot model; ][]{Lanzaetal03} and the model proposed by the Geneva team 
for the CoRoT blind test, based on a least-square fit with harmonic functions
\citep[hereinafter 200-harmonic fitting; ][]{Moutouetal05}.  Secondly, we apply the
Box-fitting Least Square algorithm \citep[or BLS; ][]{Kovacsetal02}
 to search for transits in the time series of the residuals obtained with the two different
variability models to see which one gives the best performance in terms of  
detected transits and reduced false alarms. The residuals of the transitless light curves are  analysed 
in the same way to find the frequency of false alarms  produced by  noise.

\subsection{The three-spot model of Sun-like variability}

\label{3spots}
The model is described  by \citet{Lanzaetal03}, and its performance is studied in detail
by \citet{Lanzaetal03,Lanzaetal07} in the case of the Sun at different phases of the 11-yr cycle.
Therefore, we limit ourselves to 
a brief introduction. The model makes use of three active regions,
containing both cool spots and warm faculae, to fit the rotational modulation of
the TSI, and a uniformly distributed
background component to fit the 
variation of its mean level along the 11-yr cycle. The ratio of the area
of the faculae to that of the sunspots in each active region is fixed.

The rotation
period $P_{\rm rot}$ is assumed as a free parameter ranging from 23.0 to 33.5 days,
and its value is determined, together
with the other free parameters, by minimizing the $\chi^{2}$ \citep[see ][]{Lanzaetal03}. 
The inclination
of the rotation axis $i$ is set at the solar value during the fitting
process. Considering also the mean flux level, the areas and the co-ordinates of the three active regions, 
the number of free parameters is eleven. 

The timescale of evolution of the active region pattern on the Sun is significantly
shorter than the rotation period.  The
longest time interval that can be modelled with three stable active regions is
 14 days \citep[cf.][]{Lanzaetal03}.
Hence, the  time series is subdivided into subsets of
14 days, each of which is fitted with the model described above. 
The duration of a transit is at least one 
order of magnitude shorter than the timescale of variation of the active region
configuration, ensuring that the model is not significantly affected by transit dips.  
{ Small discontinuities at the level of $\sim 30-50$ ppm are sometimes present at the matching points between successive 14-d best fits. Their impact on our results is minimized by the random choice of the origin of the TSI time series  used to simulate 
stellar activity, ensuring that those features never fall at the same points of the solar variation series.} 

\begin{figure}[!t]
\includegraphics[width=9cm,height=12cm]{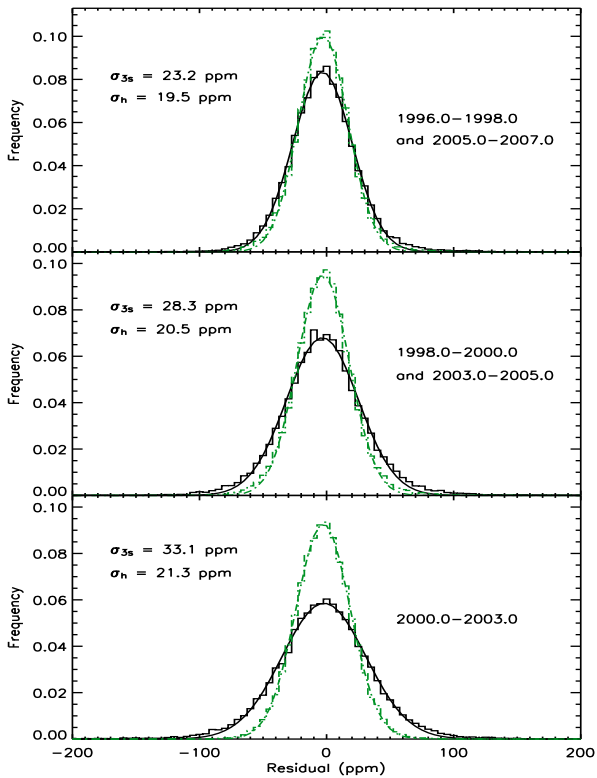} 
\caption{Distributions of the residuals of the TSI best fit along solar cycle 23 obtained with the 3-spot model
(solid black histograms) and the 200-harmonic fitting (dashed green histograms) for the three labelled time intervals
corresponding to the minimum (top panel), intermediate (middle panel) and maximum (bottom panel) of activity. 
Solid black lines and green dashed lines are Gaussian best fits to the 3-spot  and 
the 200-harmonic residual distributions, respectively. The standard deviations of the Gaussian best fits are 
indicated in the panels with $\sigma_{\rm 3s}$ and $\sigma_{\rm h}$ referring to the 3-spot model and the 
200-harmonic fitting, respectively. } 
\label{residuals}
\end{figure}

{ In Fig.~\ref{residuals}, we plot with solid lines the distributions of the residuals of the best fit to the 
TSI from 6 February 1996 to 6 February 2007, distinguishing three time intervals corresponding to the minimum,
intermediate and maximum levels of activity along solar cycle 23, respectively. The distributions are well approximated by Gaussians except in the tails that appear
slightly higher than a normal distribution. Power spectra of the residuals are plotted in Fig.~\ref{power_sp} (left panel) revealing a characteristic inverse dependence on frequency up to 0.25 hr$^{-1}$ \citep{Jenkins02} and an 
almost white spectrum
at higher frequencies.} Note the high level of power during the intermediate phase of cycle 23  
owing to the  appearance of large sunspot groups in the decaying phase of the cycle. 

\subsection{Modelling Sun-like variability by  200-harmonic fitting}
\label{harmonics}

Stellar variability is modelled as  a sum of harmonics of a fundamental frequency
$f_{\rm L} = \frac{1}{2T}$, where $T$ is the whole duration of each time series, i.e.,
$\sim 150$ days. The number of considered harmonics is fixed at $N_{\rm L}=200$, so the highest
frequency in the model corresponds to a period of 1.5 days, well below the frequency range where most 
of the transit power is concentrated. Therefore, the fitting technique is not expected to greatly affect
the transit signal.  The model flux at time
$t_{i}$ is given by: 
\begin{equation}
F_{\rm L} (t_{i}) = a_{0} + \sum_{k=1}^{N_{\rm L}} c_{k} \cos ( 2\pi k f_{\rm L} t_{i}) + 
 \sum_{k=1}^{N_{\rm L}} d_{k} \sin ( 2\pi k f_{\rm L} t_{i}). 
\end{equation}
Including the average level $a_{0}$, the model has 401 free parameters that can be 
determined by a least-square fit. We applied  Singular Value Decomposition
(hereinafter SVD) to solve the system of normal equations for this least-square problem,
as described by \citet{Pressetal92}. 

\begin{figure*}[!t]
\includegraphics[width=8cm,height=16cm,angle=90]{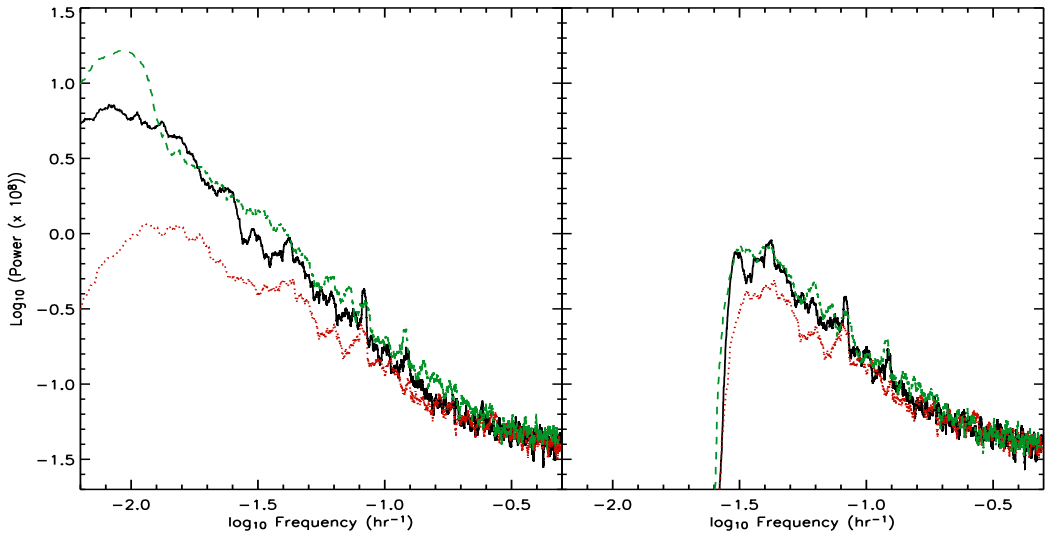}  
\caption{Power spectra of the residuals of the TSI best fit obtained with the 3-spot model (left panel) and the
200-harmonic fitting (right panel). 
Different linestyles and colors indicate different
levels of activity: black, solid line: maximum; green, dashed line: intermediate; and, red, dotted line: minimum.} 
\label{power_sp}
\end{figure*}

{ The distributions of the residuals are well approximated by Gaussians
 except in the tails that appear
slightly higher than a normal distribution (see Fig.~\ref{residuals}, dashed green plots). Their standard deviations are smaller than those of the residuals obtained with the 3-spot model. 
Power spectra of the residuals are plotted in Fig.~\ref{power_sp} (right panel) and are similar to those 
obtained with the 3-spot model, except for the
cutoff at a frequency corresponding to the period of 1.5 days. }

\subsection{Searching for planetary transits}

The time series of the residuals obtained by 
the above fitting approaches are analysed to detect 
transits by means of the BLS algorithm.
It assumes that transits are strictly periodic with a period $P$ and that the
noiseless time series assumes only two values, i.e., $H$ outside of transits and $L$ inside transits. 
In other words, each transit is assumed to have a simple box-shaped profile, that is enough for detection purposes.
The time spent in transit at the level $L$ is $qP$, where the fractionary transit length $q$ is a small
number $q \approx 0.001-0.1$. The time of the first transit in the time series is indicated by $t_{0}$.
In the presence of a white Gaussian noise with a constant standard deviation $\sigma$,
 assuming that the average level of the time series is zero, the algorithm aims to find the
best model by an estimation of the four free parameters $P$, $q$, $\delta \equiv H-L$, and $t_{0}$ that 
minimize the $\chi^{2}$ between the model and the data.   
We refer the reader to \citet{Kovacsetal02} for further details. 
Here we note only that a useful measure of the signal-to-noise ratio of a transit detection is given by:
\begin{equation}
\alpha \equiv \frac{\delta}{\sigma} \sqrt{nq},
\label{alpha_def}
\end{equation}
where $n$ is the number of points in the whole time series 
(i.e., $n \simeq 3600$ in our 150-day time series with  one-hour sampling), and $\sigma / \sqrt{nq}$ is the standard
deviation of the average of all measurements during transits. According to \citet{Kovacsetal02}, 
a transit detection can be considered 
significant at the 3-$\sigma$ level against false alarms if $\alpha \ga 6$. In the case of our tests, we  redetermine
appropriate $\alpha$ thresholds (see Sect.~\ref{alpha_deter}). 

The implementation of the BLS algorithm is performed by means of the Fortran subroutine accessible 
at {\tt http://www.konkoly.hu/staff/kovacs.html}. We slightly modified it to avoid negative values of 
$\alpha$ that would correspond to periodic positive spikes, retaining only best fits with a positive value of
$\alpha$ that correspond to transit dips. The parameter space is searched for transits with a
period $P$ between 1 and 150 days,  with 30,000 values of 
the frequency uniformly spaced between $\frac{1}{150}$ and $1$ day$^{-1}$; the fractionary length of the transit $q$ 
is varied between 0.002 and 0.08, assuming a number of 700 phase bins in the light curve folded at each trial period.
Such a sampling of the  parameter space ensures that none of the simulated transits will get lost because of
an inappropriate choice of the free parameters in our search 
\citep[for an optimal choice of the free parameter steps and ranges 
see, e.g., ][]{Kovacsetal02,Doyleetal00,Jenkinsetal02}.  

The large computational load of our experiment is managed by running our analyses on
 a large grid infrastructure consisting of three high-performance computing nodes
located in Catania, Palermo and Messina. The total number of CPUs is 560, each one being 
an AMD Opteron dual-core 64-bit processor with 1 or 2 GB of RAM \citep{Becciani07}. 
{ The CPU time for fitting solar variability  and performing transit search is on the average 
$\sim 11$ minutes per light curve with the 3-spot method and $\sim 4$ minutes per light curve with the 200-harmonic method, while about 2.5 days of elapsed time have been necessary to analyse our 
complete set of 8,000 light curves.}

\section{Results}
\label{alpha_deter}

\begin{figure}[t]
\includegraphics[width=11cm,height=9cm,angle=90]{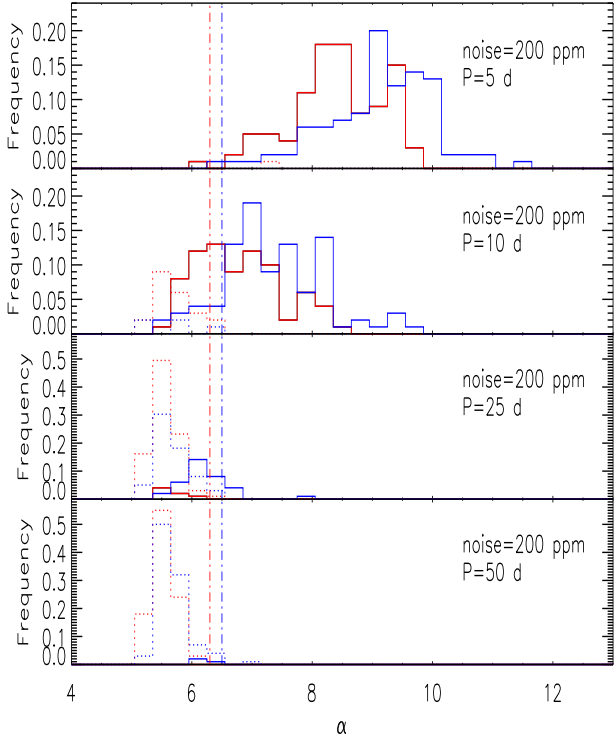} 
\caption{Distributions of the values of $\alpha$ as defined by Eq.(\ref{alpha_def}) obtained 
by analysing  light curves with transits of a planet of $1.5$ R$_{\oplus}$. Each set is 
characterized by a different orbital period of the planet (as labelled). The standard deviation of the white 
Gaussian noise is in all cases 200 ppm. { The red and blue vertical dot-dashed lines indicate 
the 1 per cent threshold for the 200-harmonic fitting and the 3-spot model, respectively.  Red solid histograms show the statistics of light curves where the period was correctly identified after applying  
the 200-harmonic method, blue solid histograms those after applying the 3-spot method, respectively.
Dashed histograms refer to the statistics of light curves where the period was incorrectly identified, with the same meaning of the colors.}}
\label{alpha_hist1.5}
\end{figure}

A typical sample of our results is given in Fig.~\ref{alpha_hist1.5}, where 
the distributions of the values of $\alpha$ are plotted for  simulations with a planet of $1.5$ R$_{\oplus}$, 
 the labelled values of the orbital period $P$, and a standard
deviation of the noise $\sigma= 200$ ppm. 
{ The vertical dash-dotted lines indicate the false-alarm thresholds  for the 200-harmonic fitting $\alpha_{\rm t} = 6.3$ and  for the 3-spot model $\alpha_{\rm t}=6.5$, respectively.  
They are derived from the analysis of the transitless  
light curves under the requisite that the frequency of false alarms be 
$\leq 0.01$ (i.e, not exceeding one in a hundred).

The histograms in Fig.~\ref{alpha_hist1.5} show the distributions of $\alpha$ for the cases where the orbital period of the planet was correctly identified or incorrectly identified, respectively. For a correct identification of the period, we require that 
the orbital period given by the BLS routine is within $\pm 0.1$ days from the true period. 
When the period is correctly identified and  $\alpha \geq \alpha_{\rm t}$, we count the detection as a positive one. 
Note that very few cases
with $\alpha > \alpha_{\rm t}$ and $P$ outside of the acceptable range are visible in the panels of
Fig.~\ref{alpha_hist1.5}. These are not counted as positive detections. }

{ The frequencies of positive detections obtained for different values of planetary radius, 
orbital period and noise level are reported in Table~\ref{det_freq}, omitting  records for values of $\sigma$ giving no detections, except for the cases with the smallest  orbital periods. 
In the first column we list the planetary radius $R$, in the second the
standard deviation of the Gaussian photon shot noise $\sigma$, in the third the orbital period $P$, in the 
fourth the frequency of positive detections with the 200-harmonic method $F_{\rm h}$ and in the fifth the 
frequency of positive detections
with the 3-spot method $F_{\rm 3s}$, respectively. 
We note that 
for a noise standard deviation of 100 ppm, the 200-harmonic
method is better than the 3-spot method, especially for planetary radii $R \leq 1.5$ R$_{\oplus}$. On the other hand, 
for higher noise levels ($\sigma \geq 200$~ppm), the 3-spot method gives a better performance than the 200-harmonic
method, especially for planetary radii $R \geq 1.5$ R$_{\oplus}$.  

As an illustrative application of our results, let us 
consider, e.g.,  a G2V star of $V=12$ observed in the CoRoT white-light exochannel with an 
integration time of one hour. In this case, $\sigma \simeq 100$ ppm and 
planets with radii as small as 
$1.25$ R$_{\oplus}$ and orbital period $P \leq 10$ days should be detectable in more than 
95 per cent of the cases if stellar variability 
and photon shot noise are the only effects limiting photometric precision. However, a proper assessment of the
performance of a given space instrument should consider also other effects 
\citep[see][]{Bordeetal03,Moutouetal05}, which goes beyond the purpose of the present study aimed 
 at comparing 
different stellar activity fitting methods. }

\begin{table}
\caption{Fraction of positive detections obtained in our experiment.}
\begin{minipage}[t]{4cm}
\setlength{\tabcolsep}{1.3mm}
\begin{tabular}{rrrcc|}
 \hline
 & & & & \\
$R\mbox{~~}$ & $\sigma\mbox{~~~}$ & $P$ & $F_{\rm h}$ & $F_{\rm 3s}$ \\
  ($R_{\oplus}$) & (ppm) & (d) & & \\ 
& & & & \\
\hline
 & & & & \\
1.0 & 100 & 5.0 & 0.93 & 0.36     \\
    &      & 10.0 & 0.31 & 0.03 \\
    &      & 25.0 & 0.01 & 0.00  \\
    &      & 50.0 & 0.00 & 0.00  \\
1.0 & 200  & 5.0 & 0.02 & 0.06   \\
    &      & 10.0 & 0.00 & 0.00  \\    
 & & & & \\
1.25 & 100 & 5.0  & 1.00  & 1.00 \\
    &      & 10.0 & 0.98  & 0.77 \\
    &      & 25.0 & 0.34  & 0.01 \\
    &      & 50.0 & 0.01  & 0.01 \\
1.25 & 200 & 5.0  & 0.46  & 0.61 \\
    &      & 10.0 & 0.06  & 0.12 \\
    &      & 25.0 & 0.00  & 0.02 \\
    &      & 50.0 & 0.00  & 0.00 \\
1.25 & 300 & 5.0  & 0.02  & 0.06 \\
    &      & 10.0 & 0.00  & 0.01 \\
    &      & 25.0 & 0.00  & 0.00 \\ 
 & & & & \\
1.5 & 100 & 5.0  &  1.00 &  1.00 \\
    &      & 10.0 & 1.00  & 1.00 \\
    &      & 25.0 & 0.97  & 0.88  \\
    &      & 50.0 & 0.16  & 0.06  \\
1.5 & 200 & 5.0  &  0.99  & 1.00  \\
    &      & 10.0 & 0.62  & 0.85  \\
    &      & 25.0 & 0.00  & 0.10  \\
    &      & 50.0 & 0.00  & 0.01  \\
1.5 & 300 & 5.0  &  0.49  & 0.61  \\
    &      & 10.0 & 0.02  & 0.17  \\
    &      & 25.0 & 0.00  & 0.00  \\      
 & & & & \\
\hline
\end{tabular}
\end{minipage}
\addtocounter{table}{-1}
\hspace*{1mm} 
\begin{minipage}[t]{4cm}
\setlength{\tabcolsep}{1.3mm}
\begin{tabular}{|rrrcc}
\hline
& & & & \\
$R\mbox{~~~}$ & $\sigma\mbox{~~~}$ & $P$ & $F_{\rm h}$ & $F_{\rm 3s}$ \\
  ($R_{\oplus}$) &  (ppm) & (d) & & \\ 
& & & & \\
\hline
 & & & & \\
1.75 & 100 & 5.0  &  1.00 & 1.00 \\
    &      & 10.0 &  1.00 &  1.00 \\
    &      & 25.0 &  1.00 &  1.00 \\
    &      & 50.0 &  0.88 & 0.77  \\
1.75 & 200 & 5.0  &  1.00 &  1.00 \\
    &      & 10.0 &  0.97 &  0.99 \\
    &      & 25.0 &  0.27 &  0.60 \\
    &      & 50.0 &  0.00 &  0.10 \\
1.75 & 300 & 5.0  &  0.95 &  0.99 \\
    &      & 10.0 &  0.36 &  0.58 \\
    &      & 25.0 &  0.00 &  0.06 \\
    &      & 50.0 &  0.00 &  0.00 \\
 & & & & \\
2.0 & 100 & 5.0  &  1.00  & 1.00   \\
    &      & 10.0 & 1.00  & 1.00   \\
    &      & 25.0 & 1.00  & 1.00   \\
    &      & 50.0 & 1.00  & 1.00   \\
2.0 & 200 & 5.0  &  1.00  & 1.00 \\
    &      & 10.0 & 1.00  & 1.00 \\
    &      & 25.0 & 0.92  & 0.99 \\
    &      & 50.0 & 0.09  & 0.57 \\
2.0 & 300 & 5.0  &  1.00  & 1.00  \\
    &      & 10.0 & 0.97  & 1.00  \\
    &      & 25.0 & 0.08  & 0.42  \\
    &      & 50.0 & 0.00  & 0.01  \\
2.0 &1000 & 5.0  &  0.00 & 0.01   \\
    &      & 10.0 & 0.00 & 0.00 \\  
 ~   & & & & \\
\hline 
 ~   & & & & \\
 ~   & & & & \\
\label{det_freq}
\end{tabular}
\end{minipage}
\end{table}


\section{Discussion}
\label{discuss}

\begin{figure}[t]
\centerline{\includegraphics[width=8cm,height=8cm,angle=90]{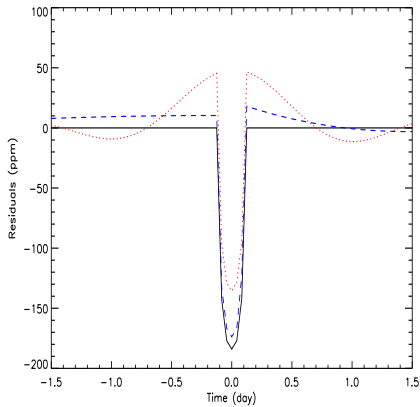}} 
\caption{Residuals obtained  with the 200-harmonic method 
(red dotted line) and the 3-spot method (blue dashed line) for a noiseless light curve containing only transits; 
the black solid line is the light variation during the transit. Note the Gibbs' phenomenon affecting the
residuals obtained with the 200-harmonic fit and the smaller distortion of the transit profile 
obtained with the 3-spot method.}
\label{gibbs}
\end{figure}

\begin{figure}[t]
\centerline{\includegraphics[width=6cm,height=6cm,angle=90]{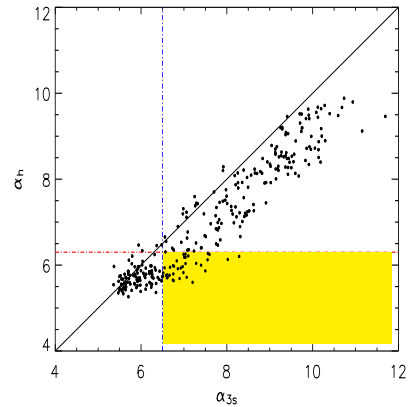}} 
\caption{Signal-to-noise ratio $\alpha_{\rm h}$ obtained with the 200-harmonic fitting  versus the corresponding value $\alpha_{\rm 3s}$ obtained with the 3-spot model for  light curves with $R = 1.5$ R$_{\oplus}$, $ P \leq 25$ d and standard deviation of the noise of 200 ppm. The dot-dashed red and blue lines indicate the $\alpha$ thresholds for the 200-harmonic fitting and the 3-spot model, respectively. The yellow region of the plot indicates transits detected with the 3-spot model that go undetected with the 200-harmonic fitting.  }
\label{scatter_plot}
\end{figure}

The results presented above prompted us to investigate the cause of the 
different performance of the two methods at different levels of noise and
planetary radii. We found that the 200-harmonic method, having 401 
degrees of freedom to fit the entire 150-d sequence, is better suited to reproduce the flux 
perturbations due to magnetic activity and noise than the 3-spot method, which has only 11 free parameters 
for each 14-d best fit { (cf.~Figs.~\ref{residuals} and \ref{power_sp})}.
However, the use of a set of orthogonal functions to fit stellar variability
produces a significant distortion of the planetary transits in the residual
light curves. In particular, the depth of the transits is reduced, as it is clear by
considering the best fits obtained in the case of a light curve containing only transits
(cf. Fig.~\ref{gibbs}). This is a consequence of the well-known Gibbs' phenomenon
\citep[e.g., ][]{MorseFeshbach54}. Since the transit dip  is mathematically 
similar to a couple of close discontinuities of the first kind, the fitting function, which is
a truncated series of orthogonal trigonometric functions, overshoots the flux variation at the 
ingress and  at the bottom of the transit leading to a reduction of its
depth $\delta$ in the residual curve. Note that the problem is not alleviated by increasing 
the order $N_{L}$ of the fitting functions, because it is a consequence of the orthogonal nature of 
the adopted set of fitting functions
\citep[cf.][]{MorseFeshbach54}. When the standard deviation of the noise is small ($\sigma=100$ ppm), the 
reduction of $\delta$  does not 
greatly affect the detection efficiency because $\alpha$, although reduced (cf. Eq.~\ref{alpha_def}), 
is  still above the significance threshold. 
Hence, the 200-harmonic method performs better than the 3-spot method since it fits better the light variations.
On the other hand, when the noise level increases ($\sigma \geq 200$ ppm), the reduction of $\delta$ has 
a larger impact on the $\alpha$-statistics as more and more values of $\alpha$ fall below the threshold
making the performance of the 200-harmonic method worse than that of the 3-spot method { (see~Fig.~\ref{scatter_plot})}. Note that the 3-spot method 
is based on the use of only three fitting functions that are not orthogonal,  thus it is not affected by the Gibbs' phenomenon. Moreover,
the reduction of the transit depth in the residual light curve is very small because the fitting
functions mainly follow the points outside transits, especially when the transit duration is 
short and the orbital period is long with respect to the adopted 14-d  time interval during which the
parameters controlling the shape of the fitting functions are kept fixed. 

The Gibbs' phenomenon affects also other filtering methods proposed to reduce the impact
of stellar microvariability, such as, e.g., that of \citet{Carpanoetal03}, or the Wiener-like
discrete filters by \citet{AigrainIrwin04}. 
Wavelet-based methods \citep[e.g., ][]{Jenkins02} may be useful to overcome the problems related 
to Gibbs' phenomenon, given the
non-orthogonal nature of their basis functions, but they are negatively affected by gaps in the time series. 
On the other hand, the non-linear iterative filter proposed by \citet{AigrainIrwin04} is  pratically 
insensitive to gaps or irregular sampling and does not make use of orthogonal functions. Therefore, it is worth a detailed comparison with our method and this will be the subject of a forthcoming work. 

Note that the 3-spot method performs  well in the presence of 
gaps \citep[see][]{Lanzaetal03,Lanzaetal07} because the model is naturally interpolated through them to follow the
observations. 
The main limitations of the 3-spot model are the computing time and the necessity of adjusting some of the parameters
if the star is not analogous to the Sun. The computation time is really not a problem when a parallel
 high-performance  computing system is available, as we have demonstrated through the extensive use of a 
grid-based system to perform our calculations. As a matter of fact, the problem of analysing the large number 
of light curves coming from a space experiment is a fully-parallelizable problem and the elapsed time scales with 
the inverse of the number of available CPUs.  

The problem of adjusting model parameters to analyse light curves of stars not stricly analogous to the Sun 
has been treated by \citet{Lanzaetal03,Lanzaetal04,Lanzaetal07} and will not be reconsidered here, since we 
intend to discuss it systematically in a successive paper.  

\section{Conclusions}

We have performed extensive numerical experiments to compare two different methods of fitting Sun-like
variability for the detection of Earth-like planetary transits. Our results show that the 3-spot model 
performs better than the 200-harmonic fitting when the standard deviation of the photon shot noise is $2-4$ times
greater than the central depth of the transit. 
Conversely, the 200-harmonic fitting has a better performance when the noise standard deviation is comparable with the depth of the transit. 
This behaviour has been interpreted as a consequence of the Gibbs' phenomenon and is expected to be common to all  
filtering and fitting methods based on sets of orthogonal functions. 
{ On the other hand,  the assumption of 
some general and simple physical hypotheses about the origin of Sun-like variability allows the 3-spot
method to reach a truly appealing performance, in spite of the fact that it has only eleven degrees of freedom.}  

Therefore, it is worth to conduct  further detailed comparisons,   in particular with  non-linear 
iterative filters, such as that proposed by \citet{AigrainIrwin04}. 
We plan to perform such tests by simultaneously applying different methods to the same, large set of simulated 
light curves, thus extending the Monte Carlo approach of the present paper. 

\begin{acknowledgements}
The authors are grateful to the Referee, Dr.~S.~Aigrain, for her comments on a previous version of the present paper that greatly helped to improve their work. 
They also gratefully acknowledge  Drs.~U.~Becciani, A.~Costa and A.~Grillo 
and  the system managers of the Trigrid and Cometa Consortia for their technical advice and 
kind assistance during the implementation and running of our numerical experiments on grid-based 
high performance computing systems.  
ASB and AFL are also grateful to Drs. R.~Alonso, P.~Barge, P.~Bord\'e, R.~Cautain, M.~Deluil, 
A.~Leger, C.~Moutou, for useful discussions on several aspects of the present work. 
The availability of unpublished data of the VIRGO Experiment
on the cooperative ESA/NASA Mission SoHO from the VIRGO Team through PMOD/WRC, Davos,
Switzerland, is gratefully acknowledged. 
The authors gratefully acknowledge support from the Italian Space Agency (ASI) under contract  ASI/INAF I/015/07/0,
work package 3170. 
This research has made use of results produced by the PI2S2 Project managed by the Consorzio COMETA, 
a project co-funded by the Italian Ministry of University and Research (MUR) within the 
{\it Piano Operativo Nazionale "Ricerca Scientifica, Sviluppo Tecnologico, Alta Formazione" (PON 2000-2006)}. 
More information is available at http://www.pi2s2.it and http://www.consorzio-cometa.it.

Active star research and exoplanetary studies at INAF-Catania Astrophysical Observatory and the Department of Physics
and Astronomy of Catania University is funded by MUR ({\it Ministero dell'Universit\`a e Ricerca}), and by {\it Regione Siciliana}, whose financial support is gratefully
acknowledged. 
This research has made use of the ADS-CDS databases, operated at the CDS, Strasbourg, France.
\end{acknowledgements}


\end{document}